# Experimental testing of the Prandtl-Tomlinson model: Molecular origin of rotational friction

Weichao Zheng[1,a)]

[1]Physical and Theoretical Chemistry Laboratory, Department of Chemistry, University of Oxford, Oxford OX1 3QZ, United Kingdom

[a)]Author to whom correspondence should be addressed: zwhich@outlook.com

Structural superlubricity, one of the most important concepts in modern tribology, has attracted lots of interest in both fundamental research and practical applications. However, the underlying model, known as the Prandtl-Tomlinson (PT) model, is oversimplified and not for real processes, despite its prevalence in frictional and structural lubricant studies. Here, with a realistic system, cholesteric liquid crystals, confined between two atomically smooth surfaces, we measure both the surface torque during rotational friction and the molecular rotation from the commensurate to incommensurate configuration at the onset of structural lubricity. Furthermore, by changing the surface potential or the strain, the Aubry transition is confirmed. The results agree well with the description by a quasi-one-dimensional version of the PT model and provide molecular evidence for rupture nucleation during static friction. Our study bridges the gap between theories and experiments and reinforces the connection between friction and fracture.

One of the 125 questions[1] published in the *Science* journal in 2021 is how to measure interfacial phenomena at the microscale. For the interface between solids and solids, contact mechanics have been well-established. However, friction, which dates back to research five centuries ago, has not been well understood due to the multiple contacts at the nanoscale and the complicated dynamics involving physical interactions and chemical reactions[2,3].

Two significant signs of progress in the understanding of stick-slip motions have been made in the past decades. First, the experimental verification[4] of structural superlubricity with two-dimensional (2D) materials manifests the validity of the one-dimensional (1D) Prandtl-Tomlinson (PT) and Frenkel-Kontorova (FK) models[5-7] that led to the development of the superlubric theory[8-10]. Another prediction based on the FK model, i.e., the Aubry transition[11], has also been observed in 1D cold ions[12] and 2D colloidal lattices[13]. However, these two models are oversimplified and the direct experimental testing has not been accomplished for several decades[14]. Second, the paradigm using brittle fracture theory[15] was developed and experimentally confirmed to describe the onset of kinetic friction from the static friction that had previously been described with the static friction coefficient. However, the mechanism of rupture nucleation during static friction at the microscopic level is still unclear[15,16]. The above puzzles are due to the difficulties of experimental characterization at the atomic interface.

Similarly, the interface plays a big role in the viscoelastic behavior of liquid crystals. Some historical puzzles, such as the discontinuous cholesteric-nematic transition[17] and permeative flows[18], have been ascribed to the boundary conditions. Recently, we proposed[18] that the historical measurement of anomalous viscosity of cholesterics in the capillary is, in fact, the twist elasticity resulting from the confinement with a rather small Ericksen number. In other words, liquid crystals are elastic solids at small scales. Furthermore, our study[17] showed that the cholesteric slip occurs only when the critical surface torque is reached. However, it is still unclear how the slip occurs. In the present work with more experimental data, we interpret the slip behaviors at the critical surface torque under mechanical winding as the reduction of friction by rotating liquid crystal molecules from the commensurate, namely the easy axis, to the incommensurate configuration, forming structural lubricity [Fig. 1(b)]. In this real-space system, the corrugated surface potential on the muscovite mica works like a grooved surface[19,20] to align molecules. At the easy axis, molecules lie on the commensurate position of grooves to minimize the twist distortion, while under stress, molecules rotate gradually to an incommensurate position with a maximum angle $\pi/2$ to the easy axis. With the decay of anchoring strength, it is easier to reach the incommensurate configuration, undergoing the Aubry transition. On the other hand, the rotation of molecules is the sign of rupture nucleation resulting in the propagation of rupture fronts until the slip occurs. Therefore, we consolidate the paradigm that static friction is interfacial rupture[15] and is indeed dynamic[21,22]. Finally, the relationship between dislocations and cracks is discussed.

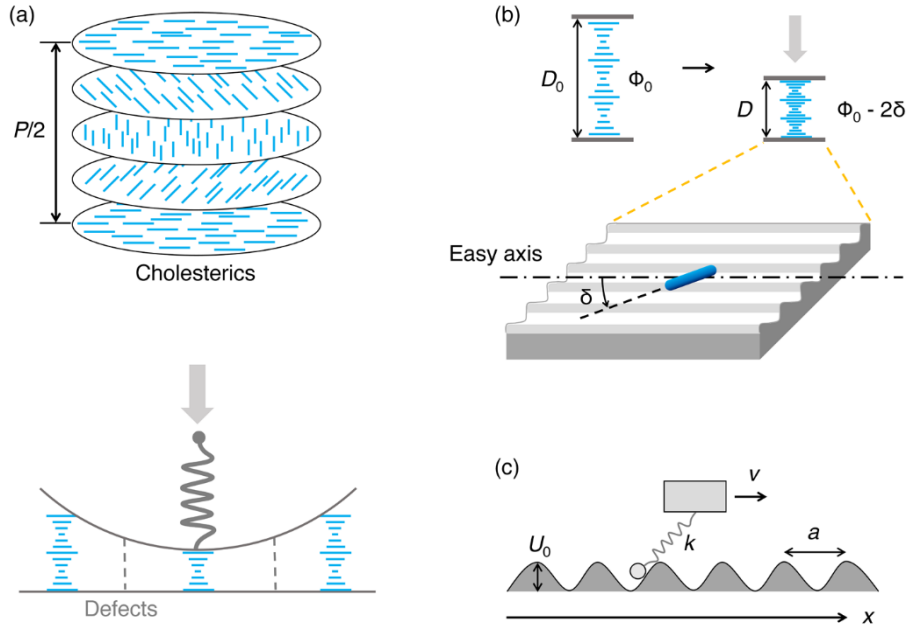

**FIG. 1.** Mechanical winding of cholesterics. (a) Schematic diagram of cholesterics under the confinement of crossed cylinders (front view), where $P = 244$ nm is the pitch. (b) The deviation of molecules from the easy axis under stress, where $\Phi_0$ is the original twist angle at the distance without compression $D_0$, and $\delta$ is the deviated angle at the compressed distance $D$. The dot-dash line is the easy axis on the potential energy surface. (c) The Prandtl-Tomlinson model, $\frac{1}{2}U_0$ and $a$ are the amplitude and periodicity of the surface potential, respectively, $k$ is the spring constant, and $v$ is the velocity along the $x$-axis.

The experiments were performed using a standard instrument[23,24], i.e., the Surface Force Balance (SFB), to measure force responses of confined cholesterics during surface approach and retraction [Fig. 1(a)]. Cholesterics are chiral liquid crystals with a pitch of about 244 nm. The main experimental procedure is to compress and stretch the confined cholesterics and measure the forces that respond. The optics generated in the SFB were recorded and analyzed to obtain results, such as distance, velocity, force, and twist angle. Detailed experimental methods are shown in Sec. SIII of the Supplementary Materials. In the SFB, the surfaces are made of muscovite mica with atomic smoothness, and the interaction between the surfaces and cholesterics is not as strong as chemical bonds. As a result, the liquid crystal molecules rotate both in the bulk and on the surface to minimize the free energy under stress [Fig.

1(b)]. The free energy per unit area $G$ as a sum of the anchoring energy (one surface) and twist elastic energy by ignoring the dislocation energy is written as[17],

$$G = \frac{1}{2}W\delta^2 + \frac{1}{2}K_{22}(\frac{\Phi}{D} - q_0)^2 D, \qquad (1)$$

where $\delta$ is the deviated angle of molecules to the easy axis, $W$ is the anchoring strength, $K_{22} = 6$ pN is the twist elastic energy, $\Phi$ is the total twist angle, $D$ is the smallest surface separation between two crossed cylinders [Fig. 1(b)], and $q_0$ is the molecular rotation rate of cholesterics at relaxation. The detailed derivations of the cholesteric equations in the following text can be found in a previous study[17]. With strong anchoring strength, twist angle $\Phi = \Phi_0 - 2\delta \approx \Phi_0 = q_0 D_0$ keeps the relaxed twist angle $\Phi_0$ at the original distance $D_0$ [Fig. 1(b)], thus,

$$G = \frac{1}{2}W(\frac{\Phi_0 - \Phi}{2})^2 + \frac{1}{2}K_{22}q_0^2\frac{(D_0 - D)^2}{D}. \qquad (2)$$

The form in Eq. (1) is widely used in the liquid crystal community, sometimes with different expressions in the anchoring energy[25,26]. In fact, this form is a modified three-dimensional (quasi-1D) version of the PT model [Fig. 1(c)] below or the FK model, with a quasi-static motion,

$$U(x,t) = \frac{1}{2}U_0 \cos(\frac{2\pi}{a}x) + \frac{1}{2}k(x - vt)^2, \qquad (3)$$

where $U$ is the free energy, $\frac{1}{2}U_0$ and $a$ are the amplitude and periodicity of the surface potential respectively, $k$ is the spring constant, $v$ is the velocity along the $x$ axis, and $t$ is the time.

The rotational friction in cholesterics under confinement is a balance of the twist elastic torque $\Gamma_e$ and the frictional anchoring torque $\Gamma_a$, by differentiating Eq. (1) with respect to the twist angle $\Phi$, while ignoring the viscous torque with a strong anchoring limit,

$$\Gamma_e = K_{22}\left(\frac{\Phi}{D} - q_0\right) = \Gamma_a = \frac{1}{2}W\frac{\Phi_0 - \Phi}{2}, \qquad (4)$$

$$K_{22}q_0\left(\frac{D_0}{D} - 1\right) = \frac{1}{2}W\frac{\Phi_0 - \Phi}{2}. \qquad (5)$$

Figure 2(a) shows a typical experiment of cholesteric slippage under mechanical winding. During surface compression at a speed smaller than 4 nm/s, cholesterics underwent an elastic response and then yielded when reaching the maximum surface torque, resulting in a surface jumping event [Fig. 2, (a) and (b)]. Figure 2(c) shows that indeed the compression ratio at the critical jumping distance remains almost constant at about 0.2-0.4, with different half-pitch layers (i.e., layer thickness = half-pitch = $\pi$ rotation) of cholesterics. If the threshold of the critical surface torque $\Gamma_c$ at the critical jumping distance $D_c$ is assumed[17],

$$D_c = \frac{D_0 - \frac{4\Gamma_c}{q_0 W}}{1 + \frac{\Gamma_c}{K_{22}q_0}}. \qquad (6)$$

From the fitting line in Fig. 2(d), the critical torque $\Gamma_c \approx 0.22$ mN/m and anchoring strength $W \approx 0.27$ mN/m can be calculated using Eq. (6) and the pitch $P = 244$ nm. Meanwhile, the deviated angle $\delta \approx 0.52\pi$ is obtained using Eq. (4), which is consistent with the previous result[17], i.e., $\delta \approx 0.49\pi$. The effect of Burgers vector and non-integer layer on the measurement is discussed in the Supplementary Materials (Fig. S1). The deviation of molecules by around $\pi/2$ from the easy axis is a sign of structural lubricity at the maximum incommensurate configuration. Therefore, the barrier to having a slip is the activation energy required to get structural lubricity. Furthermore, the critical surface torque is almost independent of the layer thickness or the motor speed of the SFB, ranging from a few nanometers per second to 80 nm/s [Fig.2, (e) and (f)].

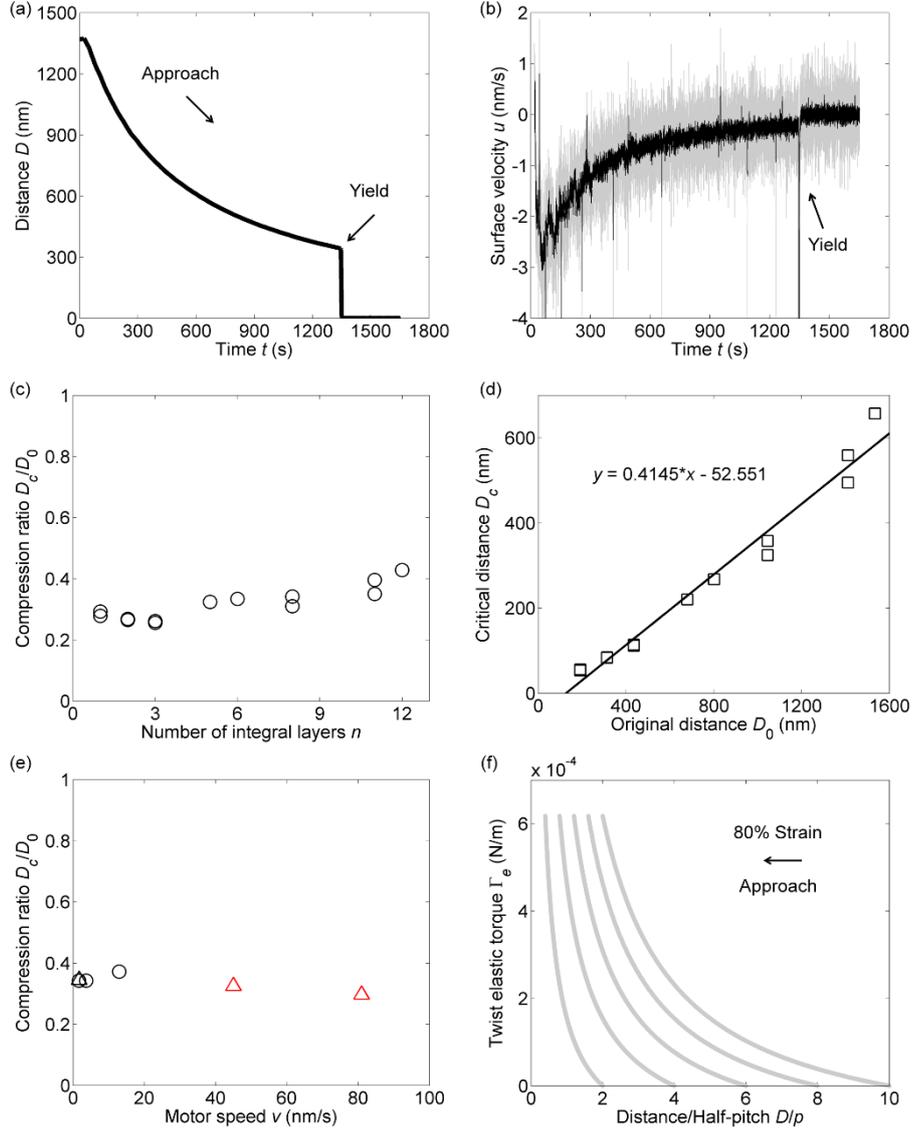

**FIG. 2.** Slippage of cholesterics at the incommensurate position. (a) The elastic response and yield of cholesterics during the surface approach experiment. (b) Corresponding surface velocity (grey). The black curve is the smoothed data. (c) The compression ratio of cholesterics with various layers in the jumping events (i.e., yield events). (d) The critical yield distance as a function of the original distance. The line is the linear fit. (e) The compression ratio of cholesterics in the jumping events with various motor speeds of the SFB. The experiments were performed with 11 (circle) and 12 (triangle) integer layers. Some data (black) were reused from a previous study[17]. (f) Twist elastic torque with various cholesteric layers under compression up to 80% strain [an arbitrary value, i.e., relative deformation $(D_0-D)/D_0$], calculated using Eq. (5).

In the PT model, a dimensionless parameter $\eta$ is calculated to predict the stick-slip behaviors, which is the ratio between the stiffness of the surface potential and the spring[3,27],

$$\eta = \frac{2\pi^2 U_0}{ka^2}. \quad (7)$$

By analogy, the parameter $\eta_f$ based on Eq. (2) is also calculated,

$$\eta_f = \frac{W}{k_{\text{chl}}}, \quad (8)$$

where $k_{\text{chl}} = K_{22}q$ is the spring constant of the cholesterics and $q = \Phi/D$.

In a previous study[17], we showed three distinct regimes of cholesterics under mechanical stress during the decay of surface anchoring (Fig. 3). First, with strong anchoring strength, cholesteric layers were collectively removed at about 65% strain, i.e., the constrained regime. Second, with intermediate anchoring strength, cholesteric layers were compressed up to about 30% strain, before being removed one by one, i.e., the stick-slip regime. Third, with weak anchoring strength, cholesteric layers were continuously removed one by one almost without strain. Although explained by the decrease of surface torque, the regime transition will be further interpreted here. With time evolution, probably due to the adsorption of water that smoothens the corrugation of the surface potential, the anchoring strength decreases. As a result, the transition goes from the constrained regime (initially $\eta_f > 1$), stick-slip, to the sliding-slip regime[17] ($\eta_f < 1$), which is the Aubry transition[14] by decreasing the surface potential (Fig. 3). In other words, the stress-induced incommensurate status changes from a pinned status to a more continuous transition (sliding-slip regime) layer by layer with a smaller corrugation of surface potential, with which molecules easier rotate to reach incommensurate status. These three regimes are consistent with the results predicted by a simulation[28]. On the other hand, the jump process (i.e., kinetic friction) is continuous in the constrained regime while it is discontinuous in the stick-slip regime (Fig. 3), which is another Aubry transition[14] by increasing the spring constant (i.e., strain-stiffening) of cholesterics with larger compression ratio[18]. Although in the constrained regime, the anchoring potential is large, after compression the stiffness of cholesterics outweighs that of the anchoring potential. With the stiffer spring constant of cholesterics, there is not enough time for the slower molecules to recover from the incommensurate configuration during slip to the commensurate configuration, such that the structural lubricity continues.

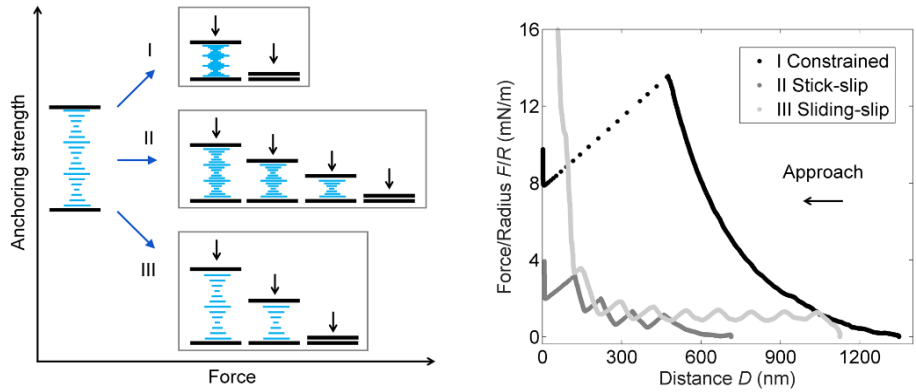

**FIG. 3.** Aubry transition by changing the anchoring strength or the stiffness of cholesterics. Three distinct regimes of force responses under compression were observed with the decay of the anchoring strength (experimental data reused)[17].

Under compression, cholesterics elastically deform and finally yield at a critical friction torque, which is also a typical fracture behavior. The rotation of molecules from the commensurate to the $0.52\pi$ incommensurate configuration during mechanical winding is a sign of rupture nucleation that cannot be observed in a solid interface with stiff materials. Thanks to the weak interaction between liquid crystal molecules and the mica surface, the anchoring deviation is quite obvious, commonly observed under a polarized microscope in the liquid crystal community[29-31]. However, without high mechanical stress, no large azimuthal deviations can be detected. With a small deviation, it is difficult to test the exact form of the anchoring energy. From what we obtained above, the parabolic potential was valid at a large deviation; however, the Rapini-Papoula potential failed[17].

With fracture mechanics, it is easier to explain the differences in the distance profile and surface velocity, and the hysteresis on force profile and twist transition during the approach and retraction of surfaces [Fig. 4, (c) to (f)]. During retraction, there was a jump-out event because of the strong adhesion at the contact position [Fig. 4, (c) to (e)]. Cholesteric layers behave like elastic solids with a small Ericksen number. The innermost dislocation defect serving as a crack breaks the neck of layers [Fig. 4(a)], which is analogous to the opening mode of fracture in a rod with cracks [Fig. 4(b)]. Simultaneously, a new layer comes to fill the increased height [Fig. 4(f)]. The neck-breaking and the slip-in are alternate layer-by-layer [Fig. 4(f)]. This breakup is not the same as the break of a liquid jet due to the Plateau-Rayleigh instability[32]. By contrast, during the approach, defects move to a larger radius, such that they never become cracks, but the whole layers undergo a tearing mode fracture encountering large energy before yielding [Fig. 4, (a) to (f)]. Under such circumstances, cholesterics behave like ceramics that resist compression but not tension.

The adhesion between two crossed cylinders or two spheres at contact during retraction has been proven to be the opening mode of fracture for decades[33]. We may also speculate that rolling friction is also the crack opening on one side that decreases the dissipated energy compared to the shearing mode of the normal stick-slip friction.

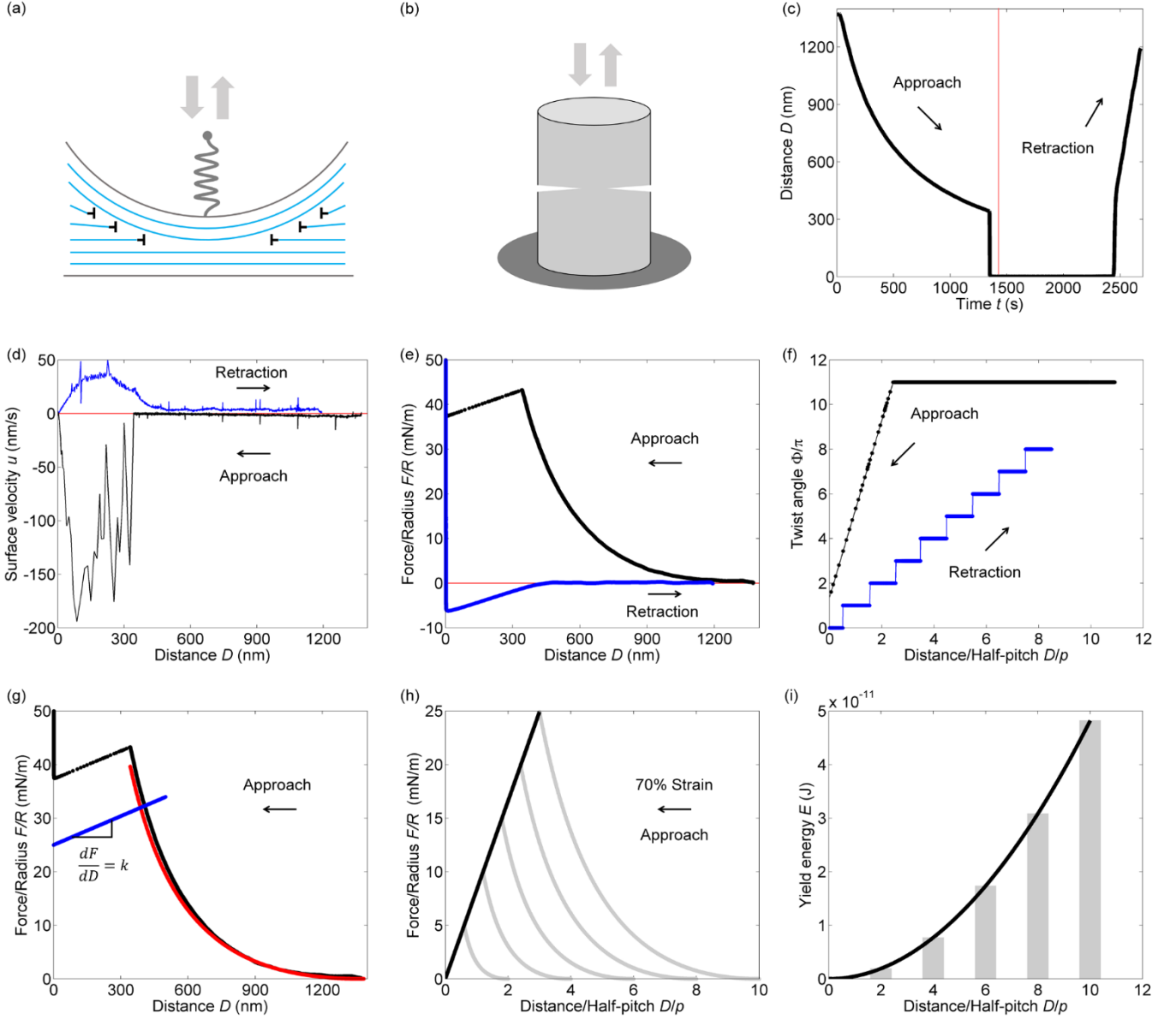

**FIG. 4.** Fractures in cholesterics. (a) Dislocations of cholesterics in the SFB. (b) A rod with two cracks. (c-f) The differences in the distance profile, surface velocity, force profile, and twist transition during approach and retraction experiments. The red line in (c-e) separates the approach and retraction regimes. The non-integer layer has been deducted in (f). (g) The force profile of cholesterics during the surface approach. The red line is the theoretical fit using Eq. (9). $R = 1$ cm is the radius of crossed cylinders. The blue line is the slope of the SFB spring with a spring constant $k = 179$ N/m. (h) The force profiles up to 70% strain (an arbitrary value) with different layers calculated by Eq. (9). The black line is the linear fit of the maximum forces. (i) The yield energy calculated by integrating the forces with respect to the distance in (h). The black line is the parabolic fit of the yield energy.

With the strong anchoring assumption, the anchoring energy is negligible compared to the twist elastic energy. Thus, the elastic force with $n$ layers is calculated by the second term on the right side of Eq. (2) with Derjaguin approximation,

$$F = 2\pi R G^n = \pi R K_{22} q_0^2 \frac{(D_0^n - D)^2}{D}. \qquad (9)$$

The measured forces can be well fitted by the force calculation using Eq. (9), confirming an elastic deformation [Fig. 4(g)]. Figure 4(h) shows that the forces generated by different layers with the same strain are different with a linear increase proportional to the number of layers, while the yield energy calculated from these force profiles increases parabolically [Fig. 4(i)], which is different from the linear increase of elastic energy in normal springs with different lengths. The increase in yield energy is consistent with a previous study[34] showing that lubricants increase the fracture energy since, in dry friction, the stiffness of the surface is quite large decreasing the stored energy.

With fracture mechanics, it is easier to understand the formation of dislocation defects in the SFB and Grandjean-Cano wedge (Fig. 5). At a height equal to $n$ integer layers, the cholesterics are relaxed. By contrast, integer cholesterics are compressed on the left side and stretched on the right side, i.e., $D_0 \pm \Delta D$ [Fig. 5(b)], which is more significant at small heights with relatively strong deformation. By increasing the heights, the layers are further stretched until a new layer is added to minimize the free energy. Therefore, with less intense deformation, the Burgers vector $b = \frac{1}{2}nP$ gradually increases at large distances[35,36]. In a more complex geometry with bumps and hollows [Fig. 5(c)], the dislocations automatically form an isoheight map. The dislocations are denser in bumps with larger heights compared to those in hollows. At large distances, thicker dislocation lines may be observed due to the increase of the Burgers vector. The geometry of confinement may provide a method to design defect networks[37] with different topological structures.

When the anchoring strength is strong, the disordered dislocations will be repelled to the bisector of the confinement, while with weak anchoring strength, the dislocations will be attracted to the surface[38]. This behavior provides a method to engineer the distribution of dislocations in materials. Furthermore, the paths of disordered dislocations are analogous to the crack paths designed on curved surfaces[39,40]. In other words, the disordered dislocations visualize the region with higher stress intensity[39,40], and the order parameter in liquid crystals serves as a single variable to predict how cracks propagate in a complex geometry. Meanwhile, the thickness of the dislocation line is associated with the levels of stress intensity.

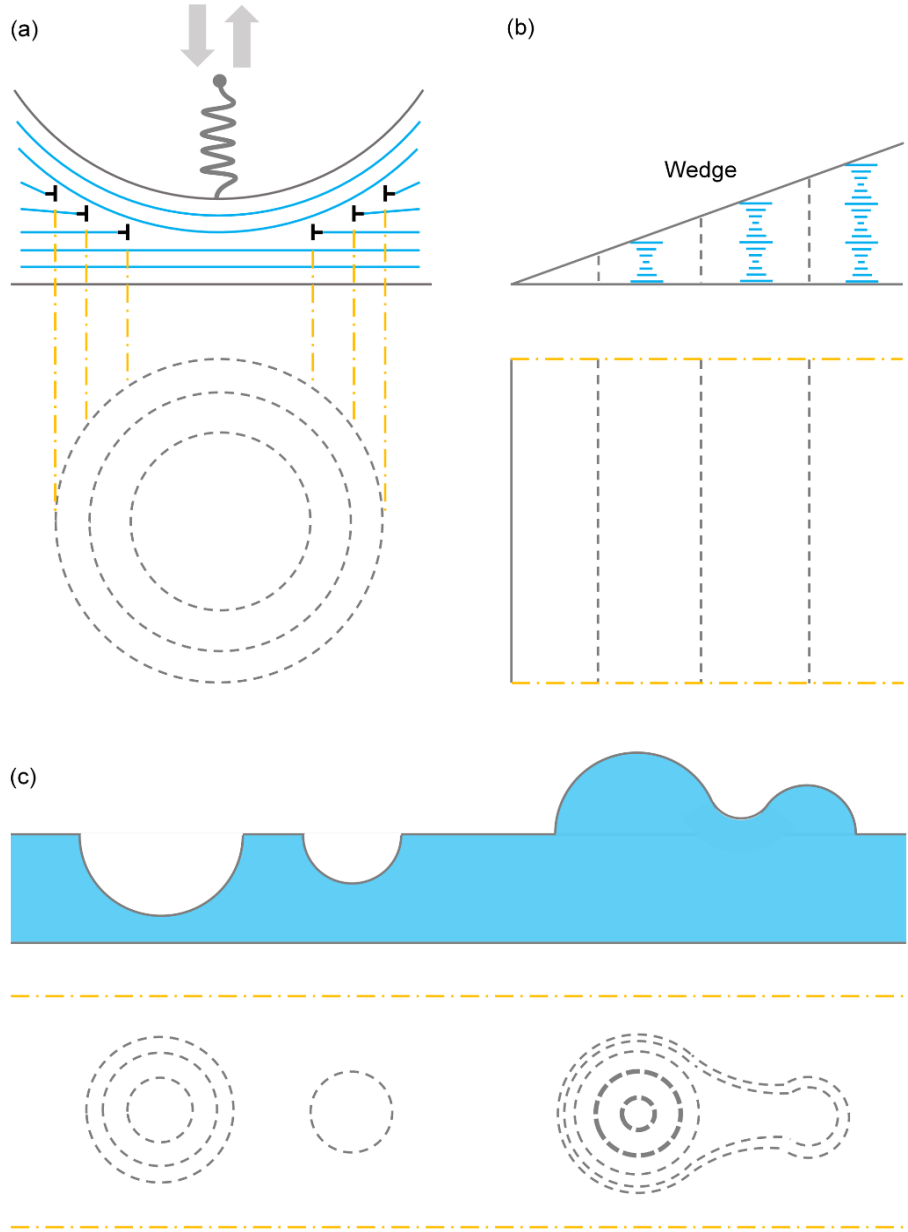

**FIG. 5.** Dislocations in different geometries. Front-view and top-view of cholesteric dislocations in (a) the SFB, (b) the Grandjean-Cano wedge, and (c) a geometry with bumps and hollows. The dashed lines represent dislocations, the dot-dashed lines are guidelines, and the thick dashed lines in (c) are thick dislocations.

Rotational friction on surfaces at the molecular level has been studied with absorbing molecules. The rotation of a single molecule can be controlled by external stimuli, such as electrons[41], temperature[42], and mechanical stress[43]. By contrast, the dynamics of absorbing molecules can be probed collectively as surface viscosity[44] or solid friction torque[45]. However, previous research mainly focused on the molecule-substrate interaction, i.e., the term of corrugated surface potential in the PT model. Without studying the interaction within molecules or between molecules, i.e., the elastic term, the PT model cannot be tested. In this work, we show that the surface potential and elastic energy are coupled and, therefore, should be studied simultaneously. Benefiting from the well-defined cholesteric system, the rotational friction at the molecular level is unraveled.

If we compare the static rotational friction with the dry sliding friction ruled by the Coulomb-Amontons laws, it is apparent that, firstly, the surface torque is independent of the contact area [Eq. (4)]. Additionally, despite different interaction areas with various layers (Sec. SII in the Supplementary Materials), the critical torque is the same [Fig. 2(c)]. Secondly, the critical surface torque is independent of the approaching speed outside the viscous torque regime [Fig. 2(e)]. Thirdly, the surface torque is proportional to the deviated angle [or deviated molecular rotation rate, Eq. (4)] rather than the normal load. However, in the Coulomb-Amontons laws, the normal load also could be described by the elastic deformation of the contact area through Hooke's law, which, therefore, is proportional to the strain. In other words, the traditional sliding friction is proportional to the strain; by contrast, the rotational friction is proportional to the molecular deviated angle. Taken together, we conclude that static rotational friction is still governed by the Coulomb-Amontons laws.

Many concepts in liquid crystals are borrowed from the crystalline community, such as dislocation and piezoelectricity. Conversely, the concept of disclination, which was developed in liquid crystals, has been used later in crystals[46]. Despite the well-developed elastic energy theory, liquid crystals are considered as fluids[47]. Therefore, fracture mechanics has rarely been used to describe liquid crystal behaviors. Even the concept of surface torque is rarely used in the liquid crystal community[17]. Our previous study[18] highlighted the importance of elasticity with strong surface anchoring under confinement. This work further underlines the friction and fractures in liquid crystals, which promotes the understanding of structural lubricity and interfacial ruptures in crystals.

Although the exact form of the anchoring potential is not proven, it is reasonable to use the parabolic potential[25], since the elastic energy of liquid crystals is based on Hooke's law. The twist interaction of molecules with the surface is similar to the elastic interaction between cholesteric molecules. In fact, the anchoring potential could be verified by measuring both the surface separation and the deviated angle, simultaneously. For example, the newly developed μSFA[48] can be mounted into the polarized microscope for this purpose. The further understanding of surface anchoring may be assisted by first-principles calculations[49,50].

It has been shown[51] that the maximum dissipated energy during friction is equal to the corrugation of the potential energy surface, which seems to be the first term of Equation 1 if the second term did not dissipate as phonons. For cholesterics, the rotation of molecules along the helical axis also causes piezoelectricity[52], which needs to be taken into account as electronic dissipation. In the stick-slip regime, apart from the dissipated heat, residual energy is also stored in the rest layers until all the layers are squeezed out.

There has been a lasting debate[53,54] on the mechanism of stick-slip motions during molecular friction. Three scenarios, including one-layer or whole-film melting and interlayer slip, have been proposed[53]. It is difficult to distinguish the melting and the slip when the heat dissipation from the yield energy is high enough to melt molecular layers or cause wear during stick-slip. However, from the results of this work, it may be more favorable for the scenario of interlayer slip[54,55], which is also a sign of interfacial ruptures at the molecular level.

In summary, we tested the PT and FK models with cholesterics under mechanical winding in the SFB. During rotational friction, governed by the Coulomb-Amontons laws, liquid crystal molecules gradually rotated from the commensurate to incommensurate configuration with a balance of the twist elastic torque and the surface torque. With strong anchoring, molecules parabolically rotated $0.52\pi$ reaching the maximum incommensurate status and structural lubricity occurred. This experiment provides a model system for studying friction with tunable surface potential, elastic constants, as well as cholesteric pitch length by external stimuli, such as light, temperature, stress, dopants, and magnetic and electric fields. It may shed light on the understanding of surface anchoring, structural lubricity, shear thickening[56], glass transition, yield stress materials, and earthquakes.

**Supplementary Material**

See the supplementary material for information about the Burgers vector, interaction areas in the SFB, and experimental methods.

W.Z. is very grateful to S. Perkin who suggested that the π/2 deviation of anchoring may be intrinsic. Some parts of this work have been discussed in the Ph.D. thesis titled "Optical and mechanical responses of liquid crystals under confinement (2020)". This research was supported by the European Research Council (Grant Nos. ERC-2015-StG-676861 and 674979-NANOTRANS).

AUTHOR DECLARATIONS

Conflict of Interest

The author has no conflicts to disclose.

Author Contributions

Weichao Zheng: Conceptualization; Data curation; Formal analysis; Methodology; Visualization; Writing – original draft.

DATA AVAILABILITY

The data that support the findings of this study are available from the corresponding authors upon reasonable request.

# Supplementary Material

# Experimental testing of the Prandtl-Tomlinson model: Molecular origin of rotational friction

Weichao Zheng[1]

[1]Physical and Theoretical Chemistry Laboratory, Department of Chemistry, University of Oxford, Oxford OX1 3QZ, United Kingdom

**I. Burgers vector and non-integer layer**

On the second day of the experiment, the sample was still in the constrained regime (Fig. S1). Notably, the data fall on two trend lines [Fig. S1(a)], indicating that the Burgers vector may affect the counting of the sample layers[1]. The two trend lines are almost parallel with similar slopes. At small distances, the Burgers vector is 1, and the dislocation defects lie in the bisector between two surfaces with strong anchoring strength[2]. With increasing distances, the Burgers vector increases and the defects stay in random positions[3], which may affect the layer transitions. By fitting the data of thin samples [Fig. S1(b)], the critical torque of 0.18 mN/m, the anchoring strength of 0.26 mN/m, and the deviated angle of $0.44\pi$ were calculated, which is consistent with the data obtained from the fresh sample. After 24 hours, the anchoring strength was almost the same, but the deviated angle decreased by a small amount. The difference in the deviated angle is reasonable, considering the systematic errors.

By contrast, the trend line for the thick samples results in a similar critical torque of 0.20 mN/m, but a strange anchoring strength of -0.03 mN/m and a deviated angle of $-4.0\pi$. The deviated angle manifests the change in the Burgers vector in thick samples. Therefore, thin samples are recommended for the precision of the measurement and more experiments are needed to test the effect of the Burgers vector. From the results, it seems reasonable to use samples with no more than 12 integer layers.

Usually, the easy axes on mica surfaces are mismatched; therefore, there is always a non-integer layer with a rotation angle less than $\pi$. This last layer does not show a jump event. With an integer layer, the total deviated angle on both surfaces is $\pi$ to reach structural lubricity. For a non-integer layer, for example, $0.8\pi$, when the molecules totally deviate towards $0.8\pi$, the distance should be close to zero to get a large elastic torque that exceeds the critical torque. However, from the trend line in Fig. 2(d), when the critical distance is zero, the original distance would be 126.8 nm, which is larger than an integer layer. Furthermore, it is impossible to squeeze a thick layer into a distance smaller than a few molecule-layer sizes. Under such circumstances, other mechanisms may dominate. For example, the molecular layers can be squeezed out one by one generating structural forces with a simple liquid[4].

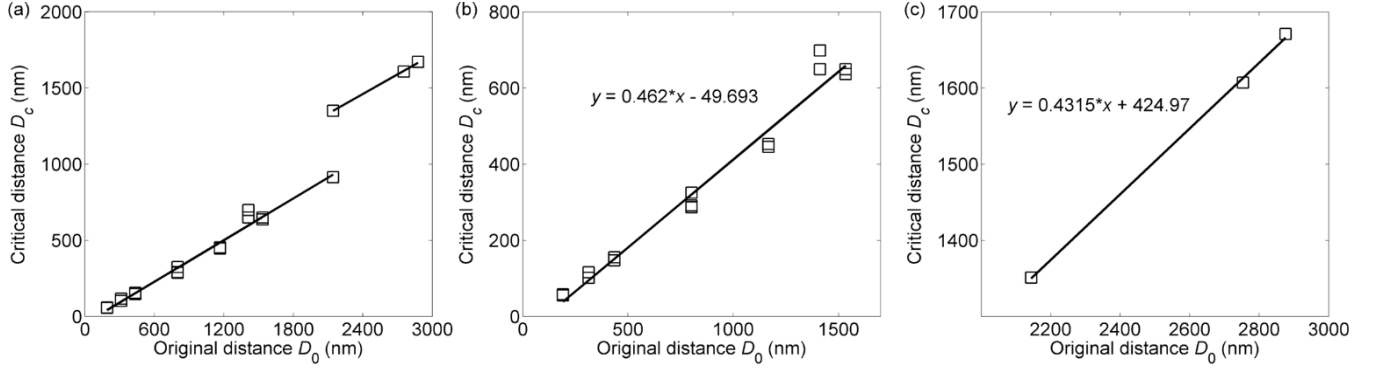

**FIG. S1.** The critical yield distance as a function of the original distance in the sample after 24 h. (a) The yield events with all sample layers. (b) The yield events with sample layers no larger than 12. (c) The yield events with thick samples.

## II. Interaction areas with various layers

With $n$ layers confined in the SFB, the original distance is $D_0^n$. In the region with more than $n$ layers, for example, $n+1$ layers, the layers are half compressed and half stretched [Fig. 1(a)], which neutralizes the interaction. In other words, the innermost layer region is the effective interaction area, although the whole sample is under confinement. When the surface separation is compressed to $D$, ignoring the dislocation energy, there is an equilibrium position[5] at the radius $r = \sqrt{2R(D_0^n - D)}$, where $R$ is the radius of the cylinder and $r$ is the distance away from the surface contact point. In other words, the region within radius $r$ is the effective interaction area. Therefore, with various layers but the same compression ratio at the jumping events, the interaction area depends on $D_0^n$, which varies with layers.

## III. Methods

The cholesterics were mixed with 62.4 wt% nematics (QYPDLC-036, similar to BL036 from Merck) and a chiral dopant (R2011, right-handed), which were purchased from Qingdao QY Liquid Crystal Co., Ltd (Chengyang, Qingdao, China). The cholesterics produce a pitch of about 244 nm.

The cholesterics were dried in the Schlenk line at around 80°C overnight for the measurement of the compression ratio. Without the Schlenk line, the strong anchoring strength may be obtained [Fig. 2, (a) and (b) and Fig. 4], but sometimes the trace amount of water in the sample may affect the experiments. To obtain a surface with atomic smoothness, muscovite mica was freshly cleaved and deposited with a silver layer on one face. During the experimental setup, the silver side of the mica was glued onto the glass lens with a radius of 1 cm. Subsequently, the lenses were mounted into the SFB with standard procedures[6]. Finally, the SFB chamber was dried with the nitrogen for at least 1 h before the injection of the cholesterics.

The SFB is a standard instrument invented decades ago[6,7]. Inside the SFB, the optics reflected between the silver layers generate fringes of equal chromatic order (FECO)[8,9] on the spectrometer. The FECO can be analyzed to obtain information about surface separation, according to multiple-beam interferometry[8,9]. For force measurements, a cylindrical lens is connected to a motor that moves with a constant speed of a few nanometers, while another lens sits on a spring with a known spring constant. When a repulsive force is encountered during surface approach, the surface separation moves slower than the motor speed. Therefore, the force can be calculated.

Both the distance and the surface velocity profiles [for example, Fig 2, (a) and (b)) were directly measured by the SFB.

However, the force profiles [for example, Fig. 4(e)] relied on the calibration of the motor speed of the SFB[1]. For cholesterics, the long-range elastic force may affect the precision of the calibration. The motor speed was estimated at the onset of the surface motion where cholesterics were under small strain. For example, from the surface velocity profile in Fig. 2(b), the motor speed was estimated at about 2.6 nm/s, with which the force profile was calculated [Fig. 4(g)]. Furthermore, the calculated force profile is well fitted by the theory and the slip region is parallel with the slope of the spring constant [Fig. 4(g)], which in turn validates the calibration.

The surface separation was directly measured through the optics. However, the original distance, determined by the relaxed cholesteric thickness, in Fig. 2 was calculated according to the number of non-integer and integer layers. The number of layers was rounded by the initial distance as the jump during retraction occurs at the position in the middle of two adjacent integer layers. Simultaneously, the jumping events were taken into account in the determination of the layer number. For example, if there are 10 layers of cholesterics with a relaxed distance of $10p$ under confinement, where $p$ is the half-pitch, but the initial surface separation is $10.2p$ (some tension exists), the original distance is still $10p$. But if the surface separation is $10.8p$, it is more likely to be 11 layers. Usually, the jumping events tell the accurate number of layers.